\documentclass[preprint,12pt]{elsarticle}

\usepackage{amsmath,amssymb}

\usepackage{booktabs}
\usepackage{tikz}
\usepackage{pgfplots}
\pgfplotsset{compat=1.16}
\usetikzlibrary{arrows.meta,positioning}
\usepackage{ytableau}
\usepackage[colorlinks=true,linkcolor=blue,citecolor=blue,urlcolor=blue]{hyperref}
\biboptions{sort&compress}

\newcommand{\code}[1]{\texttt{#1}}
\newcommand{\dg}{\mathrm{diag}}
\newcommand{\ket}[1]{\lvert #1\rangle}

\newcommand{\ketbra}[2]{\lvert #1\rangle\!\langle #2\rvert}
\newcommand{\braket}[2]{\langle #1 \vert #2 \rangle}
\newcommand{\me}[3]{\langle #1 \vert #2 \vert #3 \rangle}
\newcommand{\Id}{\hat{\mathbf 1}}

\journal{Nuclear Physics A}

\begin{document}

\begin{frontmatter}

\title{Algebraic coefficients of fractional parentage for three
particles in the oscillator Jacobi basis\tnoteref{ded}}
\tnotetext[ded]{Dedicated to the memory of Ramutis Kazys Kalinauskas
(d.~2025), friend and colleague, who saw it first.}

\author[ftmc]{A.~Step\v{s}ys\corref{cor1}}
\ead{augustinas.stepsys@ftmc.lt}
\cortext[cor1]{Corresponding author.}

\author[kk]{S.~Mickevi\v{c}ius}

\author[ftmc]{D.~Germanas}

\address[ftmc]{Center for Physical Sciences and Technology (FTMC),
Saul\.etekio al.~3, LT-10257 Vilnius, Lithuania}
\address[kk]{HEI Kauno kolegija, Pramon\.es ave.~20, 50468 Kaunas,
Lithuania}

\begin{abstract}
We present a closed-form construction of the orbital coefficients of
fractional parentage (CFP) of three identical particles in the
translationally invariant harmonic oscillator (HO) Jacobi basis.  These
coefficients are usually obtained by numerically diagonalizing a
permutation operator in every energy--angular-momentum block.  We show
that this diagonalization is unnecessary.  Dressing the class operator
$\hat P_{13}+\hat P_{23}$ with the single unitary
$\hat W=\hat H_1\hat\Phi$ --- a fixed complexification followed by the HO
bracket with mass parameter $d=1$ --- turns it into the label swap
times an integer, and every CFP is then the single ket
$\ket{K_{i,s}}=\sqrt2\,\hat Q_s\hat W^{\dagger}\ket{i}$ --- a
$\sqrt2$-scaled, sign-decorated column of the $d=1$ bracket restricted
to one $e_2$-parity of the labels --- whose norm,
$1+s\me{i}{\hat J}{i}$, is itself a single matrix element.  The same $\hat W$ carries label
parity into swap symmetry, $\hat W\hat Q_s\hat W^{\dagger}=\hat R_s$,
which is what makes the bookkeeping collapse.  The eigenvalues are the
exact integers $\pm2$ and $\pm1$ that label the S(3) symmetry classes.
The construction is uniform in $E$, $L$ and symmetry class, and reduces
the CFP-specific work from $O(n^3)$ to $O(n^2)$ in the block dimension;
an eigensolver remains only inside the bracket engine that supplies
$\hat H_1$, and all residual numerical error originates there.  We
implement it on top of an SU(3)-basis bracket engine and run the full
battery of checks over all 1301 blocks up to $E=50$ (912\,366 CFPs,
largest block $2907\times2907$): the S(3) class assignment is exact in
every block, and the residual numerical error --- entirely the bracket
engine's --- stays below $10^{-8}$ throughout $E\le35$ and is restored
to $10^{-22}$ or better by a quadruple-precision build beyond that.  We
further validate the construction against the independently generated
antisymmetrization data of a production three-nucleon code, including
the $^3$He ground-state energy of the full pipeline.
\end{abstract}

\begin{keyword}
coefficients of fractional parentage \sep
harmonic-oscillator brackets \sep
Talmi--Moshinsky transformation \sep
Jacobi coordinates \sep
three-body problem \sep
permutation group
\end{keyword}

\end{frontmatter}

\section{Introduction}

A nucleus is self-bound, so nothing physical may depend on where its
centre of mass sits.  A basis built from single-particle coordinates
does not know this and pays for it with spurious centre-of-mass
excitations; the cure is to build the basis on \emph{Jacobi
coordinates} from the start \cite{kramer,moshbook,ncsm-ti,fbs2024},
which describe only the internal geometry --- for three particles, the
separation of a chosen pair and the distance from that pair to the
spectator.

The price is that the Pauli principle becomes awkward.  Exchanging two
nucleons no longer just relabels basis states: it acts on the Jacobi
coordinates themselves, mixing them.  Enforcing antisymmetry therefore
means finding, inside each energy--angular-momentum block, the
combinations of basis states that transform correctly under S(3).
Those combinations are the orbital \emph{coefficients of fractional
parentage} (CFP), the three-particle case of the classical
spectroscopic construction \cite{hamermesh,kamun-cfp}, and the first
link in the antisymmetrization chain of translationally invariant
few-body methods \cite{ncsm-ti,ncsm-rev,liebig,fbs2024}: every larger
calculation inherits whatever they cost and whatever error they carry.

The standard way to find them is to build the matrix of a permutation
operator --- a single transposition $P_{23}$, or the class sum
$\hat\Lambda=\hat P_{12}+\hat P_{13}+\hat P_{23}$ of \cite{fbs2024},
or the
antisymmetrizer itself \cite{ncsm-ti,liebig} --- and hand it to a
numerical eigensolver.  Its matrix elements in the Jacobi
harmonic-oscillator (HO) basis are the general HO brackets (HOB)
\cite{talmi,moshinsky,smirnov,trlifaj,kamun-hob,su3hob}.  This works,
but it must be redone for every block, the blocks grow quickly with the
number of oscillator quanta, and an eigensolver costs $O(n^3)$ in the
block dimension $n$ and quietly adds round-off to numbers that ought to
be exact.

\begin{figure}[htb]
\centering
\begin{tikzpicture}[scale=1.5,>=Stealth]
  \draw[->,gray] (-1.35,0) -- (1.45,0) node[right,black] {$\boldsymbol\rho_1$};
  \draw[->,gray] (0,-1.35) -- (0,1.45) node[above,black] {$\boldsymbol\rho_2$};
  \draw[gray!35] (0,0) circle (1.2);
  \draw[very thick,blue!65!black] (0,-1.28) -- (0,1.28);
  \node[blue!65!black] at (0.30,1.30) {$P_{12}$};
  \draw[very thick,red!70!black]  (-1.28,-0.739) -- (1.28,0.739);
  \node[red!70!black] at (1.44,0.80) {$P_{23}$};
  \draw[very thick,green!45!black] (-1.28,0.739) -- (1.28,-0.739);
  \node[green!45!black] at (1.44,-0.80) {$P_{13}$};
  \draw[very thick,dashed,orange!85!black] (-1.28,-0.530) -- (1.28,0.530);
  \node[orange!85!black] at (1.40,0.28) {$T_1$};
  \draw[<->,gray] (0.62,0.358) arc (30:90:0.72);
  \node[gray] at (0.40,0.80) {\small $60^\circ$};
  \draw[<->,gray] (0.62,-0.358) arc (-30:30:0.72);
  \node[gray] at (0.95,-0.02) {\small $60^\circ$};
\end{tikzpicture}
\caption{The three transpositions of three identical particles, acting
in the plane of the two Jacobi coordinates.  Each is a \emph{reflection}
in a line, and the three lines sit $60^\circ$ apart --- the dihedral
picture of S(3).  This is where the number $3$ in
$2\cos[\pi(e_1-e_2)/3]$ comes from: a $60^\circ$ rotation of the plane
raises the phase of the two-oscillator state by $\pi/3$.  The dashed
line is the mirror $T_1$ of the $d=1$ oscillator bracket, which is not a
particle permutation at all; conjugating with it is what turns every
reflection into a plain relabelling, Eq.~\eqref{eq:plane}.}
\label{fig:mirrors}
\end{figure}
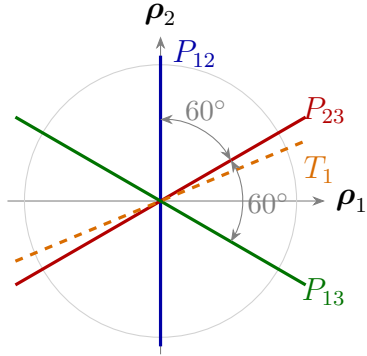

The same eigenvalue problem is met outside the oscillator Jacobi basis.
In the hyperspherical approach the symmetrized basis functions are built
recursively through ``hyperspherical coefficients of fractional
parentage'', \emph{defined} as the eigenvectors of the transposition
class sum of S($N$) and obtained by diagonalizing it
\cite{novoselsky,barnea}.  For three particles that construction is
strikingly close to the present one: the transposition acts on the
Jacobi vectors as a rotation by $\pi/3$ combined with a reflection ---
the same dihedral geometry as Fig.~\ref{fig:mirrors} below --- and the
rotation is carried out with Raynal--Revai coefficients, themselves
expressible through Talmi--Moshinsky brackets \cite{novoselsky}.  All
the ingredients of a closed form are present there too, and yet the
symmetrization is still done by a numerical eigensolver; closed
expressions in that basis have been reported only for $L=0$ and $L=1$
\cite{barnea}.  On the oscillator side, Aguilera-Navarro, Moshinsky and
Kramer \cite{amk1969} projected states of definite S(3) and S(4)
symmetry with Young operators in the single-particle basis and obtained
closed orthonormal sets for generic quantum numbers, a diagonalization
remaining necessary when all pairs $(n_sl_s)$ coincide.  The
construction below is of that kind --- explicit, and free of an
eigensolver --- but uniform: formulated directly in the Jacobi basis, it
covers every $E$, $L$ and symmetry class by one formula, and the
configurations that forced a separate treatment in \cite{amk1969} appear
here as fixed points of a single relabelling, handled by one Kronecker
delta.

This paper shows that the eigensolver is unnecessary.  \emph{The CFPs
are already tabulated inside a table few-body practitioners have been
using for sixty years} --- the harmonic-oscillator bracket itself, taken
at mass parameter $d=1$.  Up to a $\sqrt2$, a sign, and a restriction to
half of them, every CFP is one of its columns.

That such a closed form exists we first noticed empirically, in
unpublished numerical experiments.  Surrounding
$\hat\Lambda$ with
suitable diagonal phase matrices and with the $d=1$ oscillator bracket
was found to make
its matrix pseudo-antidiagonal, and the remaining rotation a sparse
matrix with entries $\pm1/\sqrt2$; merging all factors gave the
eigenvector matrix as a $\sqrt2$-scaled, sign-decorated copy of the
$d=1$ HOB itself.  The recipe was verified numerically up to $E=23$, but
the bookkeeping had been found by experiment, separately for even and
odd $E$, and no derivation was given.

In this paper we derive that result.  We work throughout inside one
fixed block and write every statement in Dirac notation, as an operator
acting on the kets of that block: the state labels then enter only
through a handful of operators that are diagonal in those kets, plus one
that merely relabels them.  The bookkeeping that had to be fixed by
experiment then --- even versus odd $E$, the central region, the phase
of the sparse rotation --- is exactly what collapses in this form, and
the central result can be read off a single ket,
Eq.~\eqref{eq:thm1ket}.  Section~\ref{sec:basis} fixes the basis and the
permutation operators.  Section~\ref{sec:swap} derives the central
identity.  Section~\ref{sec:pseudo} obtains the CFP in one formula and
reads off the symmetry classes.  Section~\ref{sec:code} describes the
implementation, its verification and its cost, and Sec.~\ref{sec:prod}
the validation against a production three-nucleon code.

\section{The chain of matrices, and what it collapses to}
\label{sec:chain}

Before any derivation it is worth seeing what the empirical recipe
actually looked like, because the whole point of what follows is that it
shrinks.

The diagonalization was originally carried out as a product of seven
factors, applied left to right to the operator to be diagonalized:

\begin{figure}[htb]
\centering
\newcommand{\chainmat}[1]{%
  \begin{tikzpicture}[scale=0.155,baseline=(current bounding box.center)]
    \foreach \c/\r in {#1}{\fill[blue!25] (\c-1,9-\r) rectangle (\c,10-\r);}
    \draw[gray!45,line width=0.12pt] (0,0) grid (9,9);
    \draw[gray!70,line width=0.3pt] (0,0) rectangle (9,9);
  \end{tikzpicture}}
\setlength{\tabcolsep}{2pt}
\begin{tabular}{ccccccccccccc}
\chainmat{1/1,1/2,2/3,2/4,3/5,3/6,4/7,4/8,5/7,5/8,6/9,7/3,7/4,8/5,8/6,9/1,9/2}
& $\times$ &
\chainmat{1/1,1/2,1/3,1/4,1/5,1/6,1/7,1/8,1/9,2/1,2/2,2/3,2/4,2/5,2/7,2/8,2/9,%
3/1,3/2,3/3,3/4,3/5,3/7,3/8,3/9,4/1,4/2,4/3,4/4,4/5,4/6,4/7,4/8,4/9,%
5/1,5/2,5/3,5/4,5/5,5/6,5/7,5/8,5/9,6/1,6/4,6/5,6/6,6/9,%
7/1,7/2,7/3,7/4,7/5,7/7,7/8,7/9,8/1,8/2,8/3,8/4,8/5,8/7,8/8,8/9,%
9/1,9/2,9/3,9/4,9/5,9/6,9/7,9/8,9/9}
& $\times$ &
\chainmat{1/1,2/2,3/3,4/4,5/5,6/6,7/7,8/8,9/9}
& $\times$ &
\chainmat{1/1,1/4,1/5,1/6,1/9,2/2,2/3,2/7,2/8,3/2,3/3,3/7,3/8,%
4/1,4/4,4/5,4/6,4/9,5/1,5/4,5/5,5/6,5/9,6/1,6/4,6/5,6/6,6/9,%
7/2,7/3,7/7,7/8,8/2,8/3,8/7,8/8,9/1,9/4,9/5,9/6,9/9}
& $\times$ &
\chainmat{1/1,2/2,3/3,4/4,5/5,6/6,7/7,8/8,9/9}
& $\times$ &
\chainmat{1/1,1/2,1/3,1/4,1/5,1/6,1/7,1/8,1/9,2/1,2/2,2/3,2/4,2/5,2/7,2/8,2/9,%
3/1,3/2,3/3,3/4,3/5,3/7,3/8,3/9,4/1,4/2,4/3,4/4,4/5,4/6,4/7,4/8,4/9,%
5/1,5/2,5/3,5/4,5/5,5/6,5/7,5/8,5/9,6/1,6/4,6/5,6/6,6/9,%
7/1,7/2,7/3,7/4,7/5,7/7,7/8,7/9,8/1,8/2,8/3,8/4,8/5,8/7,8/8,8/9,%
9/1,9/2,9/3,9/4,9/5,9/6,9/7,9/8,9/9}
& $\times$ &
\chainmat{1/1,1/2,2/3,2/4,3/5,3/6,4/7,4/8,5/7,5/8,6/9,7/3,7/4,8/5,8/6,9/1,9/2}
\\[3pt]
{\scriptsize$\pm\tfrac{1}{\sqrt2}$} && {\scriptsize$\mathrm{HOB}$} &&
{\scriptsize$i^{\,e_2}$} && {\scriptsize$\hat P_{13}{+}\hat P_{23}$} &&
{\scriptsize$(-i)^{\,e_2}$} && {\scriptsize$\mathrm{HOB}$} &&
{\scriptsize$\pm\tfrac{1}{\sqrt2}$}\\[1pt]
{\scriptsize cross} && {\scriptsize $d\!=\!1$} && {\scriptsize phases} &&
{\scriptsize class sum} && {\scriptsize phases} && {\scriptsize $d\!=\!1$} &&
{\scriptsize cross}
\end{tabular}
\caption{The original recipe, drawn for the same block $E=4$, $L=2$ as
Fig.~\ref{fig:shapes}: seven matrices multiplied together to diagonalize
one $9\times9$ operator.  The top row of labels gives what actually sits
in the entries, the bottom row what each factor is for.  Reading
outwards from the middle: the class sum is
dressed by the two diagonal phase matrices $i^{\,e_2}$ and
$(-i)^{\,e_2}$, then by the $d=1$ oscillator bracket on each side, and
finally by a sparse matrix of $\pm1/\sqrt2$ entries whose only task is
to undo the antidiagonal shape the middle produces.  Everything in this
paper says that the five inner factors collapse to
Fig.~\ref{fig:shapes}(b) and that the two outer ones are then
unnecessary.}
\label{fig:chain}
\end{figure}
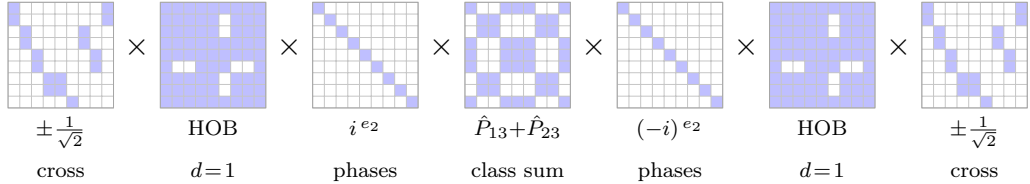

\noindent
Figure~\ref{fig:chain} shows those seven factors as they actually look,
for the same block used throughout this section.  They are of three
kinds:

\begin{itemize}\setlength{\itemsep}{2pt}
\item the two \emph{phase matrices}, diagonal, with entries
  $i^{\,e_2}$ and $(-i)^{\,e_2}$ --- the complexity they introduce
  cancels again by the end; these are $\hat\Phi$ and
  $\hat\Phi^{\dagger}$ below;
\item the two copies of the oscillator bracket at $d=1$, the dense
  factors, $\hat H_1$ below;
\item the \emph{cross matrix}, the sparse one with entries
  $\pm1/\sqrt2$, whose only job is to undo the antidiagonal shape that
  the middle of the chain produces.
\end{itemize}

\noindent
Two separate bookkeeping schemes had to be carried alongside it --- one
sorting the states by the parity of $e_2$, the other pairing each state
with its label-swapped partner --- together with a renumbering matrix
translating between them, a phase that had to be searched for
numerically, and a separate treatment of the ``central region''
$e_1=e_2$, $l_1=l_2$ that exists only for even $E$.

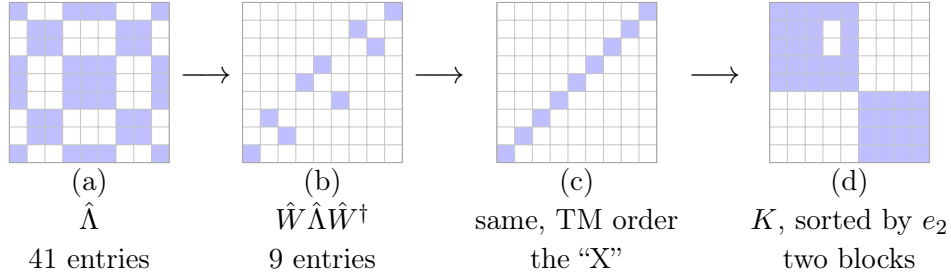
\begin{figure}[htb]
\centering
\newcommand{\shapeN}[1]{%
  \begin{tikzpicture}[scale=0.235,baseline=(current bounding box.center)]
    \foreach \c/\r in {#1}{\fill[blue!25] (\c-1,9-\r) rectangle (\c,10-\r);}
    \draw[gray!45,very thin] (0,0) grid (9,9);
    \draw[gray!70] (0,0) rectangle (9,9);
  \end{tikzpicture}}
\begin{tabular}{@{}c@{\ }c@{\ }c@{\ }c@{\ }c@{\ }c@{\ }c@{}}
\shapeN{1/1,1/4,1/5,1/6,1/9,2/2,2/3,2/7,2/8,3/2,3/3,3/7,3/8,%
        4/1,4/4,4/5,4/6,4/9,5/1,5/4,5/5,5/6,5/9,6/1,6/4,6/5,6/6,6/9,%
        7/2,7/3,7/7,7/8,8/2,8/3,8/7,8/8,9/1,9/4,9/5,9/6,9/9}
& $\longrightarrow$ &
\shapeN{1/9,2/7,3/8,4/5,5/4,6/6,7/2,8/3,9/1}
& $\longrightarrow$ &
\shapeN{1/9,2/8,3/7,4/6,5/5,6/4,7/3,8/2,9/1}
& $\longrightarrow$ &
\shapeN{1/1,1/2,1/3,1/4,1/5,2/1,2/2,2/3,2/4,2/5,3/1,3/2,3/3,3/4,3/5,%
        4/1,4/4,4/5,5/1,5/2,5/3,5/4,5/5,%
        6/6,6/7,6/8,6/9,7/6,7/7,7/8,7/9,8/6,8/7,8/8,8/9,9/6,9/7,9/8,9/9}
\\[3pt]
{\small (a)} && {\small (b)} && {\small (c)} && {\small (d)}\\
{\small $\hat\Lambda$} && {\small $\hat W\hat\Lambda\hat W^{\dagger}$}
  && {\small same, TM order} && {\small $K$, sorted by $e_2$}\\
{\small 41 entries} && {\small 9 entries} && {\small the ``X''}
  && {\small two blocks}
\end{tabular}
\caption{The actual matrices of the block $E=4$, $L=2$ ($n=9$); shaded
cells are non-zero, and nothing here is schematic.
\emph{(a)}~The class operator $\hat\Lambda=\hat P_{13}+\hat P_{23}$ in
the Jacobi basis: 41 of the 81 entries are non-zero, and this is the
matrix a numerical eigensolver is normally handed.
\emph{(b)}~The same operator dressed with $\hat W=\hat H_1\hat\Phi$.
Only nine entries survive --- one per row --- because
Eq.~\eqref{eq:thm1} makes it $\sigma\hat J\hat Z$, a permutation of the
labels times an integer.  Here the swap pairs $1\!\leftrightarrow\!9$,
$2\!\leftrightarrow\!7$, $3\!\leftrightarrow\!8$,
$4\!\leftrightarrow\!5$ and fixes $6$, so in the natural ordering it is
a permutation matrix rather than a tidy antidiagonal.
\emph{(c)}~Relabelling the states so that swapped partners sit
symmetrically about the centre --- the ``TM numeration'' of the earlier
notes --- turns the same matrix into the literal antidiagonal, running
from the top-right corner to the bottom-left, with the main diagonal
entirely empty.  This is the only thing that numeration was ever for.
\emph{(d)}~The table of coefficients $K$, with rows and columns sorted
by the parity of $e_2$ (the ``NN numeration''), splits into a
$5\times5$ and a $4\times4$ block: each coefficient lives on one parity
class and is exactly zero on the other.}
\label{fig:shapes}
\end{figure}

Everything in this paper is a statement about how that chain collapses.
It collapses in three steps, and it is worth naming them in advance:

\begin{enumerate}\setlength{\itemsep}{3pt}
\item \textbf{The middle becomes trivial.}  The four factors
  $\hat H_1\hat\Phi(\hat P_{13}{+}\hat P_{23})\hat\Phi^{\dagger}\hat H_1$
  do nothing but exchange each state with its
  label-swapped partner and multiply by an integer: panel (a) of
  Fig.~\ref{fig:shapes} becomes panel (b), one entry per row.  Ordering
  the states so that partners face each other makes it the literal
  ``X'' of the notes, panel (c).  This is Eq.~\eqref{eq:thm1} below.
\item \textbf{The cross disappears.}  The $d=1$ bracket already carries
  the symmetry that the cross was there to supply, so the two merge.
  This is Eq.~\eqref{eq:intertwine}.
\item \textbf{The two numerations become one.}  Sorting by $e_2$-parity
  and pairing by label swap are the \emph{same} operation seen from the
  two sides of the bracket, which is why a renumbering matrix was needed
  to go between them and why it is not needed now.
\end{enumerate}

\noindent
What survives is a single column of the $d=1$ bracket with a sign and a
$\sqrt2$ attached.  Sorted by the parity of $e_2$, the whole table of
coefficients falls apart into two blocks, panel (d) of
Fig.~\ref{fig:shapes}: each row is one coefficient of fractional
parentage, and each row lives on one parity block only, so that half of
every row is exactly zero --- by construction, not by cancellation.

\noindent
The rest of this paper derives that statement and checks it.

\section{The Jacobi basis and the permutation operators}
\label{sec:basis}

For three equal-mass particles we take the normalized Jacobi coordinates
\begin{equation}
  \boldsymbol\rho_1=\tfrac{1}{\sqrt2}\,(\mathbf r_1-\mathbf r_2),\qquad
  \boldsymbol\rho_2=\tfrac{1}{\sqrt6}\,(\mathbf r_1+\mathbf r_2-2\mathbf r_3),
  \label{eq:jacobi}
\end{equation}

and the translationally invariant HO basis $\ket{e_1l_1,e_2l_2:LM}$,
where $e_i=2n_i+l_i$ are the quanta of $\boldsymbol\rho_i$ and $l_1,l_2$
are coupled to the total orbital momentum $L$.  We consider one block at
a time, that is $E=e_1+e_2$ and $L$ fixed, and note the parity
constraints $(-1)^{e_i}=(-1)^{l_i}$ and $(-1)^{l_1+l_2}=(-1)^{E}$.

Let the block have dimension $n$ and abbreviate its basis kets by a
single label,
\begin{equation}
  \ket{i}\equiv\ket{e_1^{(i)}l_1^{(i)},\,e_2^{(i)}l_2^{(i)}:L},
  \qquad \braket{j}{i}=\delta_{ji},
  \qquad \Id=\sum_{i=1}^{n}\ketbra{i}{i},
  \label{eq:basis}
\end{equation}
so that $\Id$ is the identity on the block.  Everything below is an
operator built from these kets; we write operators with a hat, and
reserve unhatted matrices for the $2\times2$ transformations of the
\emph{coordinate} plane introduced in Eq.~\eqref{eq:mirror}, which live
on a different space entirely.

In this notation the object to be computed is stated at once.  A state
$\ket{K}$ of definite permutational symmetry is expanded in the
non-symmetrized basis, and the expansion coefficient
\begin{equation}
  \braket{j}{K}
  =\bigl\langle e_1^{(j)}l_1^{(j)},e_2^{(j)}l_2^{(j)}:L
   \,\big|\,K\bigr\rangle
  \label{eq:cfpdef}
\end{equation}
is the coefficient with which the parent configuration $\ket{j}$ --- a
definite pair state $(e_1^{(j)}l_1^{(j)})$ times a definite spectator
state $(e_2^{(j)}l_2^{(j)})$ --- enters $\ket{K}$.  That overlap is what
a coefficient of fractional parentage has always meant, and it is the
number this paper gives in closed form.

Let $i\mapsto\bar\imath$ be the label swap
$(e_1l_1,e_2l_2)\mapsto(e_2l_2,e_1l_1)$, an involution of the label set
whose fixed points $\bar\imath=i$, that is $e_1=e_2$ and $l_1=l_2$, are
the central states and exist only for even $E$.  Besides the scalar
\begin{equation}
  \sigma=(-1)^{E-L}\in\{\pm1\},
  \label{eq:sigma}
\end{equation}
which is constant over the block, the whole paper needs only the six
operators of Table~\ref{tab:ops}, together with
\begin{equation}
  z_i=2\cos\frac{\pi\bigl(e_1^{(i)}-e_2^{(i)}\bigr)}{3}\in\{\pm2,\pm1\},
  \label{eq:zi}
\end{equation}
an integer because $2\cos(\pi k/3)=2,1,-1,-2,-1,1$ for
$k\equiv0,\dots,5\ (\mathrm{mod}\ 6)$.  Read on a basis ket these are as
simple as an operator can be: five merely attach a number, and the sixth
relabels.  We also need the two families of projectors
\begin{equation}
  \hat Q_s=\tfrac12\bigl(\Id+s\,\sigma\hat\Pi_2\bigr),\qquad
  \hat R_s=\tfrac12\bigl(\Id+s\hat J\bigr),\qquad s=\pm1 ,
  \label{eq:proj}
\end{equation}
which select, respectively, one $e_2$-parity of the labels and one
symmetry under the swap, as the last two rows of the table record.

\begin{table}[htb]
\centering
\caption{The operators on one $(E,L)$ block used throughout.  All are
diagonal in the basis kets except $\hat J$, which merely relabels; $E$,
$L$ and $\sigma=(-1)^{E-L}$ are fixed within a block.}
\label{tab:ops}
\small
\begin{tabular}{lll}
\toprule
operator & spectral form & action on $\ket{i}$\\
\midrule
$\hat J$ & $\sum_i\ketbra{\bar\imath}{i}$ &
  $\ket{\bar\imath}$,\quad $\hat J^2=\Id$\\
$\hat\Pi_1$ & $\sum_i(-1)^{l_1^{(i)}}\ketbra{i}{i}$ & $(-1)^{l_1}\ket{i}$\\
$\hat\Pi_2$ & $\sum_i(-1)^{l_2^{(i)}}\ketbra{i}{i}$ & $(-1)^{l_2}\ket{i}$\\
$\hat\Phi$ & $\sum_i i^{\,e_2^{(i)}}\ketbra{i}{i}$ & $i^{\,e_2}\ket{i}$\\
$\hat\Psi$ & $\sum_i(-1)^{\lfloor e_2^{(i)}/2\rfloor}\ketbra{i}{i}$ &
  $(-1)^{\lfloor e_2/2\rfloor}\ket{i}$\\
$\hat Z$ & $\sum_i z_i\ketbra{i}{i}$ & $z_i\ket{i}$, Eq.~\eqref{eq:zi}\\
\midrule
$\hat Q_s$ & $\tfrac12(\Id+s\sigma\hat\Pi_2)$ &
  $\ket{i}$ if $(-1)^{e_2}=s\sigma$, else $0$\\
$\hat R_s$ & $\tfrac12(\Id+s\hat J)$ &
  $\tfrac12(\ket{i}+s\ket{\bar\imath})$\\
\bottomrule
\end{tabular}
\end{table}
All six operators of Table~\ref{tab:ops} are self-adjoint except
$\hat\Phi$, which is unitary with
$\hat\Phi^{\dagger}=\sum_i(-i)^{e_2^{(i)}}\ketbra{i}{i}$; by the parity
constraint $(-1)^{l_2}=(-1)^{e_2}$ we may use either exponent in
$\hat\Pi_2$.  We shall use $\hat\Pi_1\hat\Pi_2=(-1)^{E}\Id$,
$\hat\Phi^2=\hat\Pi_2$, $\hat\Pi_2\hat Q_s=s\sigma\hat Q_s$ and
$\hat J\hat R_s=s\hat R_s$, all of which follow by acting on $\ket{i}$.
Two further relations are used constantly.  The phase operator splits
into a sign operator times one constant phase per parity sector,
\begin{equation}
  \hat\Phi=\hat\Psi\,
   \frac{(\Id+\hat\Pi_2)+i\,(\Id-\hat\Pi_2)}{2},
  \label{eq:phipsi}
\end{equation}
and $\hat Z$ commutes with $\hat J$, because $z_{\bar\imath}=z_i$:
$2\cos[\pi(e_1-e_2)/3]$ is even under $e_1\leftrightarrow e_2$, so
\begin{equation}
  \hat Z\hat J=\hat J\hat Z .
  \label{eq:zj}
\end{equation}

On the pair $(\boldsymbol\rho_1,\boldsymbol\rho_2)$ the transpositions
act as orthogonal reflections.  In the mirror parametrization
\begin{equation}
  S(\alpha)=\begin{pmatrix}\sin\alpha&\cos\alpha\\
                           \cos\alpha&-\sin\alpha\end{pmatrix},
  \qquad S(\alpha)^2=\mathbf 1,
  \label{eq:mirror}
\end{equation}
the convention \eqref{eq:jacobi} gives
\begin{equation}
  P_{12}\to\dg(-1,1),\qquad
  P_{23}\to S\!\left(\tfrac{\pi}{6}\right),\qquad
  P_{13}\to S\!\left(\tfrac{5\pi}{6}\right).
  \label{eq:p2313}
\end{equation}

Every orthogonal or unitary $2\times2$ coordinate transformation $O$
lifts to a unitary operator $\Gamma(O)$ on the two-oscillator Fock
space,
$\Gamma(O)\,a^\dagger_i\,\Gamma(O)^{-1}=\sum_j a^\dagger_j O_{ji}$,
acting inside each block, and $\Gamma$ is a homomorphism,
$\Gamma(O_1O_2)=\Gamma(O_1)\Gamma(O_2)$.  The lift of the one-parameter
family of Talmi--Moshinsky reflections
\begin{equation}
  T(d)=\begin{pmatrix}
    \sqrt{\tfrac{d}{1+d}} & \sqrt{\tfrac{1}{1+d}}\\[4pt]
    \sqrt{\tfrac{1}{1+d}} & -\sqrt{\tfrac{d}{1+d}}
  \end{pmatrix},
  \qquad T(d)^2=\mathbf 1,
  \label{eq:Td}
\end{equation}
is, by definition, the operator whose matrix elements are the general HO
brackets for mass parameter $d$
\cite{talmi,moshinsky,smirnov,trlifaj,kamun-hob},
\begin{equation}
  \hat H_d\equiv\Gamma\bigl(T(d)\bigr),\qquad
  \me{j}{\hat H_d}{i}
   =\me{e_1l_1,e_2l_2:L}{\Gamma(T(d))}{e_1'l_1',e_2'l_2':L},
  \label{eq:hob}
\end{equation}
the bra label $j$ carrying the unprimed quantum numbers and the ket
label $i$ the primed ones.  We compute these in the SU(3) basis of
Elliott \cite{elliott} with the
engine of Ref.~\cite{su3hob}.  Being the lift of a real
reflection, $\hat H_d$ is self-adjoint, real in this basis, and an
involution: $\hat H_d^{\dagger}=\hat H_d$ and $\hat H_d^2=\Id$.
Comparing the upper-left entry of \eqref{eq:p2313} with that of
\eqref{eq:Td}, $\sqrt{d/(1+d)}=\tfrac12$ gives $d=\tfrac13$, so the
three transpositions of the block are the matrix products
\begin{equation}
  \hat P_{12}=\hat\Pi_1,\qquad
  \hat P_{23}=\hat H_{1/3},\qquad
  \hat P_{13}=\hat\Pi_2\,\hat H_{1/3}\,\hat\Pi_2 ,
  \label{eq:ops}
\end{equation}
the conjugation by $\Pi_2$ in $P_{13}$ following from
\begin{equation}
  S\!\left(\tfrac{5\pi}{6}\right)
   =\dg(1,-1)\,S\!\left(\tfrac{\pi}{6}\right)\dg(1,-1),
  \qquad \Gamma\bigl(\dg(1,-1)\bigr)=\hat\Pi_2 ,
  \label{eq:p13conj}
\end{equation}
the latter being the spatial inversion of $\boldsymbol\rho_2$.  The
S(3) relations $\hat P_{23}^2=\Id$ and
$(\hat P_{12}\hat P_{23})^3=\Id$ fix \eqref{eq:ops} completely,
including all sign conventions; we verify them numerically for all
blocks up to $E=30$ in Sec.~\ref{sec:code}.

The object to be diagonalized is the class operator restricted to the
block,
\begin{equation}
  \hat\Lambda=\hat P_{13}+\hat P_{23}
   =\hat H_{1/3}+\hat\Pi_2\hat H_{1/3}\hat\Pi_2 ,
  \label{eq:lambda}
\end{equation}
that is, twice the part of $\hat H_{1/3}$ that commutes with
$\hat\Pi_2$, which is why $\hat\Lambda$ commutes with
$\hat P_{12}=\hat\Pi_1=(-1)^{E}\hat\Pi_2$.  A remark on conventions:
\cite{fbs2024} works with the full class sum
$\hat P_{12}+\hat P_{13}+\hat P_{23}$, which differs from our
$\hat\Lambda$ by the diagonal operator $\hat\Pi_1$ and therefore has the
same eigenkets; we keep the two-term form because it is the one the
complexification of Sec.~\ref{sec:swap} acts on directly.  On the
S(3) irreps \cite{hamermesh}, using
$\hat P_{12}+\hat P_{13}+\hat P_{23}=3,0,-3$ on $[3]$, $[21]$, $[111]$,
its eigenvalues are the exact integers
\begin{equation}
  (\hat\Lambda,\hat P_{12})=(+2,+1)\ \text{on }[3],\qquad
  (-2,-1)\ \text{on }[111],\qquad
  (\mp1,\pm1)\ \text{on }[21].
  \label{eq:classes}
\end{equation}
The CFPs are the orthonormal eigenvectors, and the eigenvalue alone
already says which symmetry class a vector belongs to:
$\lambda=+2$ is the totally symmetric $[3]$,
relevant for three bosons; $\lambda=-2$ is the totally antisymmetric
$[111]$; and the two $\lambda=\mp1$ states form a mixed-symmetry
$[21]$ pair, which is the case that matters for nucleons, where the
orbital factor combines with a conjugate mixed-symmetry spin--isospin
factor to give a totally antisymmetric state \cite{kamun-cfp,fbs2024}.

\section{Why the middle of the chain becomes a swap}
\label{sec:swap}

The whole construction rests on one identity for mirrors in a single
plane, in the spirit of the complexification method used for related
Jacobi-tree transformations.

It is worth doing it once with the numbers in front of us, in the plane
itself, before lifting anything to the oscillator.  In the Jacobi plane
the two transpositions we need are the reflections
\begin{equation*}
  P_{13}\to\begin{pmatrix}
    \tfrac12 & -\tfrac{\sqrt3}{2}\\[2pt]
    -\tfrac{\sqrt3}{2} & -\tfrac12\end{pmatrix},
  \qquad
  P_{23}\to\begin{pmatrix}
    \tfrac12 & \tfrac{\sqrt3}{2}\\[2pt]
    \tfrac{\sqrt3}{2} & -\tfrac12\end{pmatrix},
\end{equation*}
which in the mirror parametrization \eqref{eq:mirror} are
$\alpha=5\pi/6$ and $\alpha=\pi/6$.  Now conjugate with
$F=\dg(1,i)$ and the Hadamard mirror $T_1$.  Taking $P_{23}$ first,
\begin{equation*}
  F\,P_{23}\,F^{*}=
  \begin{pmatrix}\tfrac12 & -\tfrac{i\sqrt3}{2}\\[2pt]
                 \tfrac{i\sqrt3}{2} & -\tfrac12\end{pmatrix},
  \qquad
  T_1\bigl(F P_{23} F^{*}\bigr)T_1=
  \begin{pmatrix}0 & e^{\,i\pi/3}\\ e^{-i\pi/3} & 0\end{pmatrix},
\end{equation*}
and the same steps for $P_{13}$ give the same antidiagonal shape with
$e^{\mp i\pi/3}$ interchanged.  Adding the two, the phases combine into
a real number and the sum is
\begin{equation*}
  T_1F\bigl(P_{13}+P_{23}\bigr)F^{*}T_1=
  \begin{pmatrix}0 & 2\cos\tfrac{\pi}{3}\\[2pt]
                 2\cos\tfrac{\pi}{3} & 0\end{pmatrix}
  =\begin{pmatrix}0&1\\1&0\end{pmatrix}.
\end{equation*}
That is the whole idea in one line: \emph{each mirror separately becomes
a swap with a complex phase, and the class sum becomes a swap with a
real one}, because the two phases are conjugate.  Everything below is
this calculation carried up to the oscillator, where $2\cos(\pi/3)$
becomes $2\cos[\pi(e_1-e_2)/3]$ and the swap becomes the exchange of the
two Jacobi labels.

Formally, let
\begin{equation}
  F=\dg(1,i),\qquad
  T_1\equiv T(1)=\tfrac{1}{\sqrt2}
   \begin{pmatrix}1&1\\1&-1\end{pmatrix}=S\!\left(\tfrac{\pi}{4}\right).
  \label{eq:FT1}
\end{equation}
A direct computation gives
$FS(\alpha)F^{*}=\left(\begin{smallmatrix}
\sin\alpha&-i\cos\alpha\\ i\cos\alpha&-\sin\alpha\end{smallmatrix}\right)$,
and conjugating with the Hadamard mirror $T_1$ exchanges diagonal and
antidiagonal,
$T_1\left(\begin{smallmatrix}a&b\\-b&-a\end{smallmatrix}\right)T_1
=\left(\begin{smallmatrix}0&a-b\\ a+b&0\end{smallmatrix}\right)$.  With
$a=\sin\alpha$, $b=-i\cos\alpha$ and
$\sin\alpha\pm i\cos\alpha=e^{\pm i(\pi/2-\alpha)}$ this yields, for
every mirror \eqref{eq:mirror},
\begin{equation}
  T_1\,F\,S(\alpha)\,F^{*}\,T_1
  =X\,\dg\bigl(e^{-i(\pi/2-\alpha)},\,e^{\,i(\pi/2-\alpha)}\bigr),
  \qquad X=\begin{pmatrix}0&1\\1&0\end{pmatrix}.
  \label{eq:plane}
\end{equation}

Each of the three factors in \eqref{eq:plane} has an elementary lift,
and each lift is an operator on the block.  The Hadamard mirror lifts to
the $d=1$ bracket, $\Gamma(T_1)=\hat H_1$.  A diagonal $2\times2$ factor
lifts to an operator that is diagonal in the kets,
$\Gamma(\dg(a,b))=\sum_i a^{\,e_1^{(i)}}b^{\,e_2^{(i)}}\ketbra{i}{i}$,
in particular
\begin{equation}
  \Gamma(F)=\hat\Phi,\qquad \Gamma(F^{*})=\hat\Phi^{\dagger}.
  \label{eq:faz}
\end{equation}
The coordinate swap $X$ lifts, in the coupled basis, to the relabelling
operator $\hat J$ times the recoupling phase of the
$(l_1l_2)L\to(l_2l_1)L$ exchange, $(-1)^{l_1+l_2-L}=(-1)^{E-L}=\sigma$,
which is constant over the block and therefore a mere number,
\begin{equation}
  \Gamma(X)=\sigma\hat J,
  \qquad\text{i.e.}\qquad
  \Gamma(X)\ket{i}=\sigma\ket{\bar\imath}.
  \label{eq:swapop}
\end{equation}
This is the first real payoff: the recoupling phase, one of the objects
that previously had to be tracked state by state, is a single number
multiplying an operator that does nothing but exchange two labels.
Table~\ref{tab:gamma} collects the lifts used below.

\begin{table}[htb]
\centering
\caption{The lifts $\Gamma$ used in this paper.  Unhatted matrices act on
the coordinate plane $(\boldsymbol\rho_1,\boldsymbol\rho_2)$; hatted
operators act on the block, and their action on a ket is in
Table~\ref{tab:ops}.}
\label{tab:gamma}
\small
\begin{tabular}{ll@{\qquad}ll}
\toprule
$2\times2$ matrix & lift & $2\times2$ matrix & lift\\
\midrule
$T(d)$, Eq.~\eqref{eq:Td} & $\hat H_d$, Eq.~\eqref{eq:hob} &
  $F=\dg(1,i)$ & $\hat\Phi$\\
$T_1=T(1)$ & $\hat H_1$ &
  $\dg(1,-1)$ & $\hat\Pi_2$\\
$S(\pi/6)$, $S(5\pi/6)$ & $\hat P_{23}$, $\hat P_{13}$, Eq.~\eqref{eq:ops} &
  $X=\left(\begin{smallmatrix}0&1\\1&0\end{smallmatrix}\right)$
  & $\sigma\hat J$\\
\bottomrule
\end{tabular}
\end{table}

Two of these lifts occur only in the combination that dresses the class
operator, and it is worth giving that combination a name:
\begin{equation}
  \hat W\equiv\hat H_1\hat\Phi,\qquad
  \hat W^{\dagger}=\hat\Phi^{\dagger}\hat H_1,\qquad
  \hat W\hat W^{\dagger}=\hat W^{\dagger}\hat W=\Id ,
  \label{eq:Wdef}
\end{equation}
unitary because $\hat H_1$ is a real orthogonal involution and
$\hat\Phi$ is diagonal and unimodular.  Everything below is a statement
about this one operator.

We can now lift \eqref{eq:plane} factor by factor.  For a single mirror,
writing $w=e^{\,i(\pi/2-\alpha)}$,
\begin{equation}
  \hat H_1\hat\Phi\,\Gamma(S(\alpha))\,\hat\Phi^{\dagger}\hat H_1
  =\Gamma\bigl(X\dg(\bar w,w)\bigr)
  =\sigma\,\hat J\hat\Omega_\alpha,\qquad
  \hat\Omega_\alpha=\sum_i\bar w^{\,e_1^{(i)}}w^{\,e_2^{(i)}}\ketbra{i}{i}.
  \label{eq:onemirror}
\end{equation}
For $\hat P_{23}$ we have $\alpha=\pi/6$ and $w=e^{i\pi/3}$, so that
\begin{equation}
  \hat\Omega_{\pi/6}\equiv\hat\Omega
   =\sum_i e^{-i\pi\left(e_1^{(i)}-e_2^{(i)}\right)/3}\ketbra{i}{i},
  \label{eq:Omega}
\end{equation}
while for $\hat P_{13}$ we have $\alpha=5\pi/6$ and $w=e^{-i\pi/3}$, so
that $\hat\Omega_{5\pi/6}=\hat\Omega^{\dagger}$.  Adding the two mirrors
gives $\hat\Omega+\hat\Omega^{\dagger}=\hat Z$, and with \eqref{eq:zj}
we obtain the central identity of this paper: in every block
\begin{equation}
  \boxed{\;
  \hat W\,\hat\Lambda\,\hat W^{\dagger}
   =\sigma\,\hat J\hat Z=\sigma\,\hat Z\hat J ,
  \qquad\text{that is}\qquad
  \hat W\,\hat\Lambda\,\hat W^{\dagger}\ket{i}=\sigma z_i\,\hat J\ket{i}
  \;} .
  \label{eq:thm1}
\end{equation}
Written out in full labels, using Table~\ref{tab:ops}, this is
\begin{equation}
  \hat H_1\,\hat\Phi\,\hat\Lambda\,\hat\Phi^{\dagger}\,\hat H_1
  \ket{e_1l_1,e_2l_2:L}
  =(-1)^{E-L}\,2\cos\frac{\pi(e_1-e_2)}{3}\;
  \ket{e_2l_2,e_1l_1:L} .
  \label{eq:thm1ket}
\end{equation}

Equation~\eqref{eq:thm1} is the whole content of the paper in one line:
\emph{dress the class operator with $\hat W$, apply it to any basis
state, and you get back that same state with its two Jacobi labels
interchanged, multiplied by an integer.}  Nothing is mixed and nothing
summed --- one state in, one state out --- which is the
pseudo-antidiagonal form found in the numerical experiments, and also
the form in which the result is easiest to check.  That $\hat Z$ is
self-adjoint with real eigenvalues means the complexity introduced by
$\hat\Phi$ has cancelled; ordering the states so that swapped partners
are adjacent makes $\hat J$ block-antidiagonal and displays the ``X''
shape explicitly.

Two immediate consequences are worth recording.  Since
$(\sigma\hat J\hat Z)^2=\hat Z^2$ and $\hat Z^2\ket{i}=z_i^2\ket{i}$
with $z_i^2\in\{4,1\}$,
\begin{equation}
  \hat\Lambda^2
   =\hat W^{\dagger}\hat Z^2\hat W,
  \qquad
  \bigl(\hat\Lambda^2-\Id\bigr)
  \bigl(\hat\Lambda^2-4\,\Id\bigr)=0 ,
  \label{eq:minpoly}
\end{equation}
so the S(3) class relation \eqref{eq:classes} is a consequence of
\eqref{eq:thm1} rather than an input.  And \eqref{eq:onemirror} shows
that a single transposition has the same swap structure but with complex
eigenvalues,
\begin{equation}
  \hat W\hat P_{23}\hat W^{\dagger}=\sigma\hat J\hat\Omega ;
  \label{eq:P23swap}
\end{equation}
the class sum is real precisely because
$\hat\Omega+\hat\Omega^{\dagger}$ is, which is why it is the more
convenient object.  Equation~\eqref{eq:P23swap} is used in
Sec.~\ref{sec:pseudo} to show that the two states of a $[21]$ pair close
under S(3).

\section{The coefficient itself, and the class it carries}
\label{sec:pseudo}

The operator $\sigma\hat J\hat Z$ of \eqref{eq:thm1} is diagonalized by
the eigenprojectors $\hat R_s$ of $\hat J$: since $\hat Z$ commutes with
$\hat J$ and hence with $\hat R_s$,
\begin{equation}
  (\sigma\hat J\hat Z)\,\hat R_s=s\,\sigma\hat Z\,\hat R_s,
  \qquad s=\pm1 .
  \label{eq:JZproj}
\end{equation}
Concretely $\hat R_s$ builds the symmetric or antisymmetric combination
of a swapped pair, and this is the ``cross'' of the empirical recipe:
\begin{equation}
  \hat R_s\bigl(\ket{i}+s\ket{\bar\imath}\bigr)
    =\ket{i}+s\ket{\bar\imath},
  \qquad
  \hat R_s\ket{i}=\ket{i}\ \text{or}\ 0
  \quad(\bar\imath=i,\ s=\pm1).
  \label{eq:cross}
\end{equation}
Combining \eqref{eq:thm1} and \eqref{eq:JZproj} already diagonalizes
$\hat\Lambda$, but one more symmetry removes both $\hat R_s$ and the
complex phases.

The $2\times2$ relation $XT_1=T_1\dg(1,-1)$ lifts, by \eqref{eq:swapop}
and $\Gamma(\dg(1,-1))=\hat\Pi_2$, to
\begin{equation}
  \hat J\,\hat H_1=\sigma\,\hat H_1\,\hat\Pi_2 ,
  \label{eq:h1swap}
\end{equation}
that is, the $d=1$ bracket intertwines the label swap with the
$e_2$-parity sign.  Written on matrix elements this is a symmetry of the
bracket table itself,
\begin{equation}
  \me{\bar\imath}{\hat H_1}{j}=\sigma\,(-1)^{l_2^{(j)}}\,
  \me{i}{\hat H_1}{j},
  \label{eq:h1swapme}
\end{equation}
and in terms of the projectors \eqref{eq:proj} it becomes a plain
conjugation by the dressing operator \eqref{eq:Wdef},
\begin{equation}
  \boxed{\;\hat W\,\hat Q_s\,\hat W^{\dagger}=\hat R_s\;},
  \qquad s=\pm1 ,
  \label{eq:intertwine}
\end{equation}
since $\hat\Phi$ and $\hat Q_s$ are both diagonal and therefore commute,
leaving $\hat H_1\hat Q_s\hat H_1=\hat R_s$; equivalently
$\hat H_1\hat Q_s=\hat R_s\hat H_1$, because
$\hat H_1\hat Q_s=\tfrac12\hat H_1(\Id+s\sigma\hat\Pi_2)
=\tfrac12(\hat H_1+s\hat J\hat H_1)=\hat R_s\hat H_1$.
Equations \eqref{eq:thm1} and \eqref{eq:intertwine} together are the
paper: \emph{$\hat W$ is the one unitary that simultaneously brings the
class operator to the label swap and carries label parity into swap
symmetry.}  The second half is the whole content of the ``NN
numeration'' of the earlier experiments --- selecting one $e_2$-parity
of the labels is the same thing as selecting one swap symmetry of the
states --- and it is what makes the bookkeeping collapse.

For each label $i$ and each sign $s=\pm1$ define
\begin{equation}
  \boxed{\;
  \ket{K_{i,s}}=\sqrt2\;\hat Q_s\,\hat W^{\dagger}\ket{i}\;}
  \label{eq:Ws}
\end{equation}
\emph{This ket is the CFP}, the letter $K$ standing for
\emph{kilminiai}, the Lithuanian term for the parentage coefficients
used in the group's earlier work.  The state carries its own labels: by
\eqref{eq:WLW} and \eqref{eq:WPW} below, the pair $(i,s)$ fixes the S(3)
eigenvalues as
$(\hat\Lambda,\hat P_{12})=\bigl(s\sigma z_i,\,s(-1)^L\bigr)$, so no
post-hoc classification of the output is ever required.  Its components are read off the $d=1$
bracket table directly,
\begin{equation}
  \braket{j}{K_{i,s}}=\sqrt2\;(-1)^{\lfloor e_2^{(j)}/2\rfloor}\,
    \me{j}{\hat H_1}{i}\;
    \delta_{(-1)^{e_2^{(j)}},\,s\sigma}\, ,
  \label{eq:Wscomp}
\end{equation}
a column of the bracket table, sign-decorated and half-emptied by the
Kronecker delta.  Two remarks on \eqref{eq:Wscomp}.  It uses
\eqref{eq:phipsi} to replace $\hat\Phi^{\dagger}$ by the real sign
operator $\hat\Psi$ on the sector selected by $\hat Q_s$.  There the two
differ by one constant, $\hat\Phi^{\dagger}=c_s\hat\Psi$ with $c_s=1$
for $s\sigma=+1$ and $c_s=-i$ for $s\sigma=-1$, so that
\eqref{eq:Wscomp} is $c_s^{-1}$ times \eqref{eq:Ws} on that sector.
Since $|c_s|=1$ the factor multiplies a whole state and changes no
physical content; \emph{we absorb it once and for all and take
\eqref{eq:Wscomp} as the definition}, which is the real form the code
stores and the form in which the two equations agree.  And $\hat H_1$ is
symmetric, so ``column'' and ``row'' are
the same statement; we say column throughout, and what is actually
computed is the overlap $\braket{j}{K_{i,s}}$ of Eq.~\eqref{eq:cfpdef}.

Its properties now follow mechanically from \eqref{eq:intertwine}, with
no phases to track.  First, by \eqref{eq:h1swap} and
$\hat\Pi_2\hat Q_s=s\sigma\hat Q_s$,
\begin{equation}
  \ket{K_{\bar\imath,s}}=s\,\ket{K_{i,s}},
  \label{eq:JW}
\end{equation}
so the two labels of a swapped pair generate the same state and only one
of them need be kept.  Second, using $\hat Q_s\hat Q_{s'}
=\delta_{ss'}\hat Q_s$ and then \eqref{eq:intertwine},
\begin{equation}
  \braket{K_{i,s}}{K_{j,s'}}
   =2\,\me{i}{\hat W\hat Q_s\hat Q_{s'}\hat W^{\dagger}}{j}
   =\delta_{ss'}\,2\,\me{i}{\hat R_s}{j}
   =\delta_{ss'}\,\me{i}{\Id+s\hat J}{j} .
  \label{eq:orth}
\end{equation}
Third, inserting $\hat\Lambda=\hat W^{\dagger}(\sigma\hat J\hat Z)\hat W$
from \eqref{eq:thm1} and using
$\hat W\hat Q_s\hat W^{\dagger}=\hat R_s$, $\hat J\hat R_s=s\hat R_s$
and $\hat Z\hat R_s=\hat R_s\hat Z$, the CFP ket is an \emph{eigenket}
of the class operator,
\begin{equation}
  \hat\Lambda\ket{K_{i,s}}
   =\sqrt2\,\hat W^{\dagger}(\sigma\hat J\hat Z)\hat R_s\ket{i}
   =s\,\sigma\,z_i\ket{K_{i,s}},
  \label{eq:WLW}
\end{equation}
while $\hat\Pi_1\hat Q_s=(-1)^{E}s\sigma\hat Q_s=s(-1)^{L}\hat Q_s$
gives in the same way an eigenket of the pair transposition,
\begin{equation}
  \hat P_{12}\ket{K_{i,s}}=s\,(-1)^{L}\ket{K_{i,s}} .
  \label{eq:WPW}
\end{equation}
Setting $j=i$ in \eqref{eq:orth} gives the norm as a single formula,
with no case distinction to make,
\begin{equation}
  \braket{K_{i,s}}{K_{i,s}}=1+s\,\me{i}{\hat J}{i},
  \qquad \me{i}{\hat J}{i}=\delta_{\bar\imath i}\in\{0,1\},
  \label{eq:norm}
\end{equation}
the matrix element being $1$ exactly on the fixed points of $\hat J$.
Both of the awkward cases of the empirical recipe are contained in this
one expression: off the fixed points the ket \eqref{eq:Ws} is already
normalized, on a fixed point the $s=+1$ ket has norm $\sqrt2$ and must
be rescaled, and the $s=-1$ ket vanishes identically.  The even/odd-$E$
split of the earlier experiments is thus nothing but the presence or
absence of fixed points of $\hat J$, recorded by the single Kronecker
delta $\delta_{\bar\imath i}$.

Equations \eqref{eq:JW}--\eqref{eq:norm} are exactly the ingredients
that had been assembled by hand, and in this notation each is a single
statement about one state.  Equation~\eqref{eq:JW} is the redundancy:
the labels $i$ and $\bar\imath$ give the same CFP up to the sign $s$,
which is the pairing of the TM numeration.  Equation~\eqref{eq:norm} is
the normalization, absorbing the ``central region'' into one Kronecker
delta.  Equations \eqref{eq:WLW} and \eqref{eq:WPW} say that every state
is simultaneously an eigenket of $\hat\Lambda$ and of $\hat P_{12}$,
with eigenvalues known in advance.

The redundancy need not be described in prose either: summed over
\emph{all} labels and \emph{both} signs, with no retention rule at all,
\begin{equation}
  \sum_{i=1}^{n}\sum_{s=\pm1}\ketbra{K_{i,s}}{K_{i,s}}
  =2\sum_{s}\hat Q_s\hat W^{\dagger}
    \Bigl(\sum_i\ketbra{i}{i}\Bigr)\hat W\hat Q_s
  =2\sum_s\hat Q_s=2\,\Id ,
  \label{eq:complete}
\end{equation}
using only $\hat W^{\dagger}\hat W=\Id$ and
$\hat Q_++\hat Q_-=\Id$.  The factor $2$ \emph{is} the double counting,
displayed by the identity rather than described: each two-element orbit
of $\hat J$ is reached from both of its labels, and each fixed point
contributes a single ket of norm squared $2$.  Storing each state once
is therefore an implementation matter --- keep one label per orbit of
$\hat J$, both signs for a two-element orbit and, by \eqref{eq:norm},
only $s=+1$ rescaled by $1/\sqrt2$ for a fixed point.  That is
$2\times(\text{number of two-element orbits})+(\text{number of fixed
points})=n$ states.  Which label of an orbit is kept is a genuine
convention and not a free choice once the output is to be compared with
stored files: by \eqref{eq:JW} the other representative flips the sign
of every $s=-1$ state.  We fix it by retaining the label with
$e_1\le e_2$, and the code follows that rule.

Collecting the retained kets over both signs $s$ therefore gives a
complete orthonormal set for the block,
\begin{equation}
  \begin{gathered}
    \hat\Lambda\ket{K_{i,s}}=\lambda_{(i,s)}\ket{K_{i,s}},\qquad
    \hat P_{12}\ket{K_{i,s}}=s\,(-1)^{L}\ket{K_{i,s}},\\[2pt]
    \lambda_{(i,s)}=s\,\sigma\,
      2\cos\frac{\pi\bigl(e_1^{(i)}-e_2^{(i)}\bigr)}{3} .
  \end{gathered}
  \label{eq:eigs}
\end{equation}
This is the main result, and it is worth stating without symbols:
\emph{to get the coefficients of fractional parentage of three
particles, take the harmonic-oscillator bracket at $d=1$, keep every
second column, and attach a sign and a factor $\sqrt2$.}  The
eigenvalue labelling each state is the exact integer $z_i$ of
\eqref{eq:zi}, so it is never computed numerically at all.

The pairs $(\lambda,\hat P_{12})$ land exactly on the S(3) classes
\eqref{eq:classes}.  A fixed point of $\hat J$ has $z_i=2$ and
contributes the single state $\ket{K_{i,+}}$ with $(\pm2,\pm1)$, that is
$[3]$ for $\sigma=+1$ and $[111]$ for $\sigma=-1$.  A two-element orbit
contributes two states, and which classes they carry is decided by $z_i$
alone.

If $|z_i|=1$, equivalently $e_1-e_2\not\equiv0\pmod 3$, the two states
carry $\lambda=\mp1$.  No one-dimensional irrep of S(3) admits
$\lambda=\pm1$, so the pair must belong to a two-dimensional irrep; that
the two kets do span an S(3)-invariant subspace follows from
\eqref{eq:P23swap} in one line.  Since
$\hat W\ket{K_{i,s}}=\sqrt2\,\hat R_s\ket{i}$ by
\eqref{eq:intertwine}, and since
\begin{equation}
  \sigma\hat J\hat\Omega\,\hat R_s\ket{i}
   =\tfrac{\sigma}{2}\bigl(\omega_i\ket{\bar\imath}
     +s\,\bar\omega_i\ket{i}\bigr)
   \in\operatorname{span}\{\ket{i},\ket{\bar\imath}\},
  \qquad
  \omega_i=e^{-i\pi(e_1^{(i)}-e_2^{(i)})/3},
  \label{eq:closure}
\end{equation}
conjugating back with $\hat W^{\dagger}$ shows that $\hat P_{23}$ maps
$\operatorname{span}\{\ket{K_{i,+}},\ket{K_{i,-}}\}$ into itself.  With
$\hat P_{12}$ diagonal on the same pair by \eqref{eq:WPW}, and
$\hat P_{13}=\hat P_{12}\hat P_{23}\hat P_{12}$, the orbit carries
$[21]$.

If instead $|z_i|=2$, the two states carry $\lambda=\pm2\sigma$, and the
same invariant subspace is \emph{reducible}: it is $[3]\oplus[111]$, one
state of each, since $\lambda=\pm2$ forces
$\hat P_{13}=\hat P_{23}=\pm1$.  This is not an exceptional case --- it
occurs for every two-element orbit with $e_1\equiv e_2\pmod 3$, which
includes both $e_1=e_2$ with $l_1\neq l_2$ and $e_1-e_2\equiv3\pmod6$.
The S(3) content of a block is therefore fixed by two integers per
orbit, its size and its $z_i$:

\medskip
\begin{center}
\begin{tabular}{llll}
\toprule
orbit type & $z_i$ & states & classes produced\\
\midrule
fixed point & $+2$ & 1 &
  one $[3]$ ($\sigma=+1$) or $[111]$ ($\sigma=-1$)\\
two-element, $|z_i|=1$ & $\pm1$ & 2 & a $[21]$ pair\\
two-element, $|z_i|=2$ & $\pm2$ & 2 & one $[3]$ \emph{and} one $[111]$\\
\bottomrule
\end{tabular}
\end{center}
\medskip

\noindent
This is what ``the states arrive already labelled'' means, and it is
precisely what a general eigensolver cannot deliver: it returns an
arbitrary orthonormal basis of a hugely degenerate eigenspace, in which
the pairing of $[21]$ partners must afterwards be reconstructed.  The
S(3) content of a block is thus obtained by counting orbits of
$\hat J$, with no matrix entry needed; this is the counting used in
Sec.~\ref{sec:prod}.  For fermions the totally antisymmetric states are
assembled from these orbital classes and the conjugate spin--isospin
S(3) irreps in the standard way \cite{hamermesh,kamun-cfp,liebig}, the
present construction supplying the orbital factors in closed form.

\subsection{One block worked out in full}
\label{sec:worked}

Take $E=2$, $L=0$.  The block has $n=3$ states, which we list in the
order
\begin{equation}
  \ket{1}=\ket{0\,0,\,2\,0:0},\qquad
  \ket{2}=\ket{1\,1,\,1\,1:0},\qquad
  \ket{3}=\ket{2\,0,\,0\,0:0}.
  \label{eq:ex-basis}
\end{equation}
Everything the recipe needs can be written down by inspection.  The
swap operator acts as
\begin{equation}
  \hat J\ket{1}=\ket{3},\qquad
  \hat J\ket{3}=\ket{1},\qquad
  \hat J\ket{2}=\ket{2},
  \label{eq:ex-J}
\end{equation}
one two-element orbit and one fixed point.  Since $E-L=2$,
$\sigma=+1$.  The values of $e_2$ are $2,1,0$, so
$\hat\Pi_2\ket{1}=+\ket1$, $\hat\Pi_2\ket{2}=-\ket2$,
$\hat\Pi_2\ket{3}=+\ket3$, and the projectors of \eqref{eq:proj} keep
\begin{equation}
  \hat Q_{+}:\ \{\ket1,\ket3\},\qquad
  \hat Q_{-}:\ \{\ket2\}.
  \label{eq:ex-Q}
\end{equation}
Finally $e_1-e_2=-2,0,+2$ gives
\begin{equation}
  \hat Z\ket{1}=-\ket1,\qquad
  \hat Z\ket{2}=+2\ket2,\qquad
  \hat Z\ket{3}=-\ket3,
  \label{eq:ex-Z}
\end{equation}
because $2\cos(\mp2\pi/3)=-1$ and $2\cos 0=2$.

The eigenvalues now follow from \eqref{eq:eigs} with no arithmetic
beyond a sign.  The two-element orbit carries $z=-1$ and contributes one
state for each $s$, with $\lambda=s\sigma z=\mp1$: a mixed-symmetry
$[21]$ pair.  The fixed point carries $z=2$ and contributes the single
state $\ket{K_{2,+}}$ with $\lambda=+2$, the symmetric $[3]$ state,
while $\ket{K_{2,-}}=0$ by \eqref{eq:norm}.  There is no $[111]$ state
in this block, and indeed the totally antisymmetric three-particle state
of $E=2$ lives in the neighbouring block $L=1$, which is
one-dimensional: there $\sigma=(-1)^{2-1}=-1$ and $\hat Z=2$ on its
single (central) state, so $\lambda=-2$ immediately, with no
diagonalization whatsoever.

The states themselves are just as direct.  In the ordering
\eqref{eq:ex-basis} the matrix elements of the $d=1$ bracket are
\begin{equation}
  \me{j}{\hat H_1}{i}=\begin{pmatrix}
    \tfrac12 & -\tfrac{1}{\sqrt2} & \tfrac12\\[3pt]
   -\tfrac{1}{\sqrt2} & 0 & \tfrac{1}{\sqrt2}\\[3pt]
    \tfrac12 & \tfrac{1}{\sqrt2} & \tfrac12
  \end{pmatrix}_{ji},
  \label{eq:ex-H1}
\end{equation}
a table the reader may check is symmetric and squares to the identity,
as $\hat H_1^{\dagger}=\hat H_1$ and $\hat H_1^2=\Id$ require.  The
sign operator acts as $\hat\Psi\ket{1}=-\ket1$,
$\hat\Psi\ket{2}=+\ket2$, $\hat\Psi\ket{3}=+\ket3$, since
$\lfloor e_2/2\rfloor=1,0,0$.  Substituting into \eqref{eq:Ws} --- keep
the $j$ of the right parity, weight by the sign and by
$\me{j}{\hat H_1}{i}$, multiply by $\sqrt2$ --- gives the three CFPs
directly:
\begin{equation}
  \begin{aligned}
    \ket{K_{1,+}}&=\tfrac{1}{\sqrt2}\bigl(-\ket{1}+\ket{3}\bigr),
      &\quad \lambda&=-1, &\quad&[21],\\
    \ket{K_{1,-}}&=-\ket{2},
      &\quad \lambda&=+1, &\quad&[21],\\
    \ket{K_{2,+}}&=\tfrac{1}{\sqrt2}\bigl(\ket{1}+\ket{3}\bigr),
      &\quad \lambda&=+2, &\quad&[3],
  \end{aligned}
  \label{eq:ex-cfp}
\end{equation}
which is exactly what the code prints for \code{cfp3 2 0}.  (The third
line is the central case: \eqref{eq:Ws} delivers $\ket1+\ket3$ of norm
$\sqrt2$, and \eqref{eq:norm} instructs us to drop the $\sqrt2$.)  One
may verify the eigenvalue claim on the spot: since
$\hat P_{12}=\hat\Pi_1$ multiplies $\ket1,\ket2,\ket3$ by $+1,-1,+1$,
the first state has $\hat P_{12}\ket{K_{1,+}}=+\ket{K_{1,+}}$, matching
$s(-1)^L=+1$ from \eqref{eq:WPW}.

Note how the three structural features of Sec.~\ref{sec:pseudo} appear
in miniature: the labels $1$ and $3$ generate the same states up to sign,
$\ket{K_{3,s}}=s\ket{K_{1,s}}$, so only one of them is kept; the state
generated by the fixed point needs no $\sqrt2$; and each CFP is
supported on a single $e_2$-parity of the labels --- the $[21]$ partner
with $\lambda=+1$ lives entirely on $\ket2$, the other two entirely on
$\{\ket1,\ket3\}$.  Every CFP is thus structurally zero on one of the
two parity sectors; here those sectors have dimensions $2$ and $1$, and
only when they are of equal size is the fraction of exact zeros exactly
one half.

A useful consequence is that the S(3) content of a block becomes label
counting.  By \eqref{eq:eigs} the spectrum of $\hat\Lambda$ is read
directly
off the diagonal of $\sigma\hat Z$ and the orbit structure of
$\hat J$: each
two-element orbit contributes $\pm\sigma z_i$ and each fixed point
contributes $2\sigma$ once.  No matrix entry is needed.  The simplest
example is the block $E=2$, $L=1$, which is one-dimensional: its single
state $\ket{1\,1,1\,1:1}$ is a fixed point of $\hat J$, so $z_1=2$ and
$\sigma=-1$ give $\lambda=-2$, the lowest totally antisymmetric
three-particle state, obtained with its exact eigenvalue and no
diagonalization at all.  For fermions the totally antisymmetric states
are assembled from these orbital classes and the conjugate
spin--isospin S(3) irreps in the standard way
\cite{hamermesh,kamun-cfp,liebig}; the present construction
supplies the orbital factors in closed form.

\section{Does it work, and what does it cost?}
\label{sec:code}

We implemented \eqref{eq:Ws} in the Fortran module on top of the unmodified SU(3)-basis~\cite{elliott} HOB engine of
Ref.~\cite{su3hob}.  Building a block costs one evaluation of
$\hat H_1$ followed by the sign $\hat\Psi$, the parity mask $\hat Q_s$
and a column selection; \eqref{eq:Ws} translates into the code line for
line.  The eigenvalues are assigned as exact integers $s\sigma z_i$, and
the $d=\tfrac13$ bracket is needed only inside the verification battery.

Every check of the battery is an operator identity, tested on the whole
block as the largest deviation over all matrix elements.  For every
block we verify $\hat H_1^2=\Id$; $\hat P_{23}^2=\hat H_{1/3}^2=\Id$;
$(\hat P_{12}\hat P_{23})^3=(\hat\Pi_1\hat H_{1/3})^3=\Id$, the S(3) presentation,
which validates all sign conventions in \eqref{eq:ops};
$\braket{K_{i,s}}{K_{j,s'}}=\delta_{ij}\delta_{ss'}$ for the retained
states; $\hat\Lambda\ket{K_{i,s}}=\lambda_{(i,s)}\ket{K_{i,s}}$ with the closed-form
$\lambda$ of \eqref{eq:eigs}; and the class consistency
\eqref{eq:classes} together with the definite $\hat P_{12}$ parity of
every retained state.  The identities \eqref{eq:thm1}, \eqref{eq:h1swap} and
\eqref{eq:intertwine} are equivalent recombinations of these and were
checked separately during development.

We ran the battery in double precision over the whole range
$E\le50$: 1301 blocks, 912\,366 CFPs, the largest block being
$2907\times2907$ at $E=50$, $L=16$, in 56~min on one core of a desktop
CPU.  The class assignment \eqref{eq:classes} is correct in every one of
the 1301 blocks, without exception.  The residual errors grow with the
shell number --- in trend rather than monotonically, since the worst
block of a shell depends on which $(E,L)$ carries the largest SU(3)
multiplicities~\cite{elliott} --- from $\sim10^{-15}$ at $E=0$ to
$\sim7\times10^{-10}$ at $E=30$ and $\sim3\times10^{-6}$ at $E=50$.
This growth is floating-point accumulation in the SU(3)-tower
construction of the bracket engine \cite{su3hob}, not in the CFP recipe,
whose eigenvalues are exact integers assigned from $s\sigma z_i$.

Where that accumulation crosses a given accuracy requirement is
therefore a statement about the engine.  Against a $10^{-8}$ threshold
the default build is clean for every block with $E\le35$ (648 blocks,
worst $8.8\times10^{-9}$); the first block above threshold appears at
$E=36$, and beyond it the degradation accelerates until at $E=50$ all 51
blocks of the shell exceed it, the worst being $3.2\times10^{-6}$ at
$L=2$.  Over the full range 297 of the 1301 blocks are above threshold;
Table~\ref{tab:fail40} gives the shell-by-shell picture.  Near the onset
the affected blocks are those of small $L$ --- all eight at $E\le40$
have $L\le4$ --- where the SU(3) multiplicities, and with them the
numerical null-space construction of the engine, are largest; higher up
that description stops being useful, since from $E=49$ the whole shell
is affected and the worst block is at $L=11$.

It should be said plainly that this is a limitation of the
\emph{implementation} and not of the construction.  What
Eq.~\eqref{eq:Ws} needs is the $d=1$ bracket table; the errors above are
those of the particular engine we use to produce it, which builds the
brackets through an SU(3) tower with an internal null-space
construction.  Supplied with exact brackets, the recipe returns exact
CFPs with exact integer eigenvalues, since it performs no numerical
linear algebra of its own.  In that sense the eigensolver has been
removed from the CFP step and relocated to the bracket engine, where it
is shared with every other user of the brackets; and the accuracy
ceiling quoted here is the engine's, not the method's.  A closed-form
bracket engine would remove it altogether --- see Sec.~\ref{sec:concl}.

For such shells the code provides a quadruple-precision build, in which
every \code{real(8)} is promoted to \code{real(16)} and the internal
Jacobi-eigensolver threshold is tightened from $10^{-14}$ to $10^{-30}$
--- the same remedy adopted in the four-particle bracket codes
\cite{hotb,hotb2}.  It restores full margins everywhere we have tested it: over
the range $E\le23$ the battery errors do not exceed
$2.4\times10^{-28}$; all eight blocks above threshold at $E\le40$, rerun
individually, fall between $2.6\times10^{-27}$ and $7.5\times10^{-24}$
(five of them are the worst-per-shell entries of
Table~\ref{tab:fail40}; the remaining three are $E=37$, $L=1$ with
$1.2\times10^{-26}$, $E=39$, $L=1$ with $4.4\times10^{-24}$ and $E=40$,
$L=4$ with $1.0\times10^{-26}$);
and re-running the worst block of each
shell from $E=44$ to $E=50$ individually gives the last column of
Table~\ref{tab:fail40}, between $2.6\times10^{-25}$ and
$5.0\times10^{-23}$, that is sixteen orders of magnitude of margin
recovered at the point where double precision is weakest.  We have not
run the quadruple build over a whole shell, so the claim made here is
the one the data supports: wherever a block was identified as
above threshold and recomputed, the margin was restored.  The price is a $\sim40\times$ slowdown of
the software-emulated arithmetic, which is why the quadruple build is
applied per block rather than to the whole sweep.

\begin{table}[htb]
\centering
\caption{Degradation of the default double-precision build beyond
$E=35$, shell by shell (from the full $E\le50$ battery).  Every block
with $E\le35$ stays below the $10^{-8}$ acceptance threshold; listed are
the largest block dimension of the shell, the worst error of the shell
and the block $(L,n)$ where it occurs, and how many blocks of the shell
are at or above threshold.  Near the onset the affected blocks are those
of small $L$; by $E=49$ the whole shell is affected and the worst block
is no longer among them.  The last column is the same worst block
recomputed with the quadruple-precision build, for those we ran
individually.}
\label{tab:fail40}
\small
\setlength{\tabcolsep}{5pt}
\begin{tabular}{cccccrc}
\toprule
&&\multicolumn{3}{c}{double precision}&&quadruple\\
\cmidrule(lr){3-5}\cmidrule(lr){7-7}
$E$&$n_{\max}$&worst error&$L$&$n$&$\ge10^{-8}$&worst block\\
\midrule
36&1183&$2.0\times10^{-8}$&2&513&1/37&$3.7\times10^{-25}$\\
37&1274&$1.7\times10^{-8}$&3&684&2/37&$2.6\times10^{-27}$\\
38&1365&$5.5\times10^{-8}$&2&570&1/39&$1.3\times10^{-24}$\\
39&1470&$6.4\times10^{-8}$&3&760&2/39&$5.0\times10^{-27}$\\
40&1575&$4.1\times10^{-8}$&2&630&2/41&$7.5\times10^{-24}$\\
41&1680&$8.1\times10^{-8}$&1&462&2/41&---\\
42&1800&$6.5\times10^{-8}$&2&693&10/43&---\\
43&1920&$6.6\times10^{-8}$&1&506&15/43&---\\
44&2040&$5.1\times10^{-7}$&2&759&20/45&$7.8\times10^{-25}$\\
45&2176&$6.6\times10^{-7}$&1&552&26/45&$5.0\times10^{-23}$\\
46&2312&$4.6\times10^{-7}$&2&828&31/47&$6.6\times10^{-25}$\\
47&2448&$1.5\times10^{-6}$&1&600&42/47&$4.9\times10^{-23}$\\
48&2601&$1.9\times10^{-6}$&2&900&43/49&$1.1\times10^{-24}$\\
49&2754&$6.4\times10^{-7}$&11&2520&49/49&$2.6\times10^{-25}$\\
50&2907&$3.2\times10^{-6}$&2&975&51/51&$3.4\times10^{-24}$\\
\bottomrule
\end{tabular}
\end{table}

The practical value of \eqref{eq:Ws} is that it removes an eigensolver
from the inner loop, and we measured what this is worth.  The companion
program \code{bench.f90} builds the same $K$ of every block twice.  The
traditional route evaluates $\hat H_{1/3}$, forms
$\hat\Lambda=\hat H_{1/3}+\hat\Pi_2\hat H_{1/3}\hat\Pi_2$, splits it
into its two $\hat\Pi_1$ sectors (which is what gives the eigenvectors a
definite pair parity) and diagonalizes each with a LAPACK symmetric
eigensolver; this is $O(n^3)$.  The direct route is \eqref{eq:Ws}:
evaluate $\hat H_1$ and copy out the columns, which is $O(n^2)$.  Both
pay for the same SU(3) tower construction and one bracket evaluation, so
we time those separately in Table~\ref{tab:timing}.

Since the comparison is only as honest as the competitor is fast, we
first tuned the traditional route, running the eigensolver with all
three LAPACK drivers (\code{dsyev}, \code{dsyevd}, \code{dsyevr}) and
with both the netlib reference implementation and OpenBLAS 0.3.29.
Summed over all 1301 blocks with $E\le50$ it takes $565$, $275$, $140$
and $67$~s respectively for reference \code{dsyev}, OpenBLAS
\code{dsyev}, \code{dsyevr} and \code{dsyevd}; at $E=50$ the same four
numbers are $108.3$, $51.4$, $22.9$ and $12.2$~s.  The best
configuration, OpenBLAS with \code{dsyevd}, is thus $8.9\times$ faster at
$E=50$ than the reference \code{dsyev} against which such comparisons
are often made, and it is what we quote throughout.  Multithreading does
not help at these block sizes (sixteen threads give $21.7$~s at $E=50$
against $12.2$~s on one core), so all timings are single-core.  The
direct route calls no BLAS or LAPACK at all and is unaffected by any of
these choices.

\begin{table}[htb]
\centering
\caption{Time to build the CFP matrices of one shell, summed over all
$L$ (gfortran 14.2, \code{-O2}, OpenBLAS 0.3.29 with the
divide-and-conquer driver \code{dsyevd}, one core, three repetitions per
block).  The first three columns are the direct route \eqref{eq:Ws} ---
the shared tower construction, the shared bracket evaluation, and the
column assembly that replaces the eigensolver; \code{dsyevd} is the eigensolver
of the traditional route.  ``direct'' and ``diag.'' are the corresponding
end-to-end totals.  The last two columns are the speedups: ``step'' is
the ratio \code{dsyevd}/assembly of the CFP-specific work alone, which is
what \eqref{eq:Ws} replaces, and ``total'' is the ratio diag./direct of
the end-to-end times, diluted by the shared bracket engine.}
\label{tab:timing}
\footnotesize
\setlength{\tabcolsep}{1.2pt}
\begin{tabular}{ccccccccccc}
\toprule
&&\multicolumn{3}{c}{direct \eqref{eq:Ws}}&\multicolumn{2}{c}{diagonalization}
 &\multicolumn{2}{c}{end to end}&\multicolumn{2}{c}{speedup}\\
\cmidrule(lr){3-5}\cmidrule(lr){6-7}\cmidrule(lr){8-9}\cmidrule(lr){10-11}
$E$&$n_{\max}$&prepare&$H_1$&assembly&$H_{1/3}$&\code{dsyevd}
 &direct&diag.&step&total\\
&&[s]&[s]&[s]&[s]&[s]&[s]&[s]&&\\
\midrule
16&147&0.004&0.004&$<$0.001&0.004&0.003&0.008&0.011&---&1.4\\
24&405&0.039&0.035&0.0038&0.034&0.035&0.078&0.108&9.1&1.4\\
30&726&0.160&0.134&0.0148&0.134&0.167&0.309&0.461&11.3&1.5\\
36&1183&0.556&0.470&0.0650&0.469&0.706&1.091&1.731&10.9&1.6\\
40&1575&1.180&1.013&0.1613&1.011&1.654&2.354&3.845&10.3&1.6\\
45&2176&2.672&2.840&0.3897&2.854&4.707&5.902&10.233&12.1&1.7\\
50&2907&6.238&6.867&0.8142&6.954&12.185&13.919&25.376&15.0&1.8\\
\midrule
\multicolumn{2}{l}{all 1301 blocks, $E\le50$}
 &39.12&39.97&4.87&40.05&67.31&83.96&146.49&13.8&1.74\\
\bottomrule
\end{tabular}
\end{table}

The step that \eqref{eq:Ws} replaces behaves as expected.  Fitting a
power of $n$ over the range $E=35\dots50$ gives $n^{3.14}$ for
\code{dsyevd} against $n^{2.67}$ for the column assembly; the latter moves
$n^2$ data, and its measured exponent is inflated by the way it is timed,
as the difference of two nearly equal wall times.  At $E=50$ the
comparison is $12.2$~s of eigensolving against $0.8$~s of copying, a
factor $15$, growing slowly with $n$.  End to end the gain is much
smaller, $1.8\times$ at $E=50$ and $1.74\times$ cumulatively over all
1301 blocks with $E\le50$, with measured exponents $n^{2.82}$ against
$n^{2.97}$, because both routes pay for the bracket engine, which at
$E=50$ is $94\%$ of the cost of the direct route.  The CFP construction
has thus ceased to be the bottleneck: what remains is the cost of the HO
brackets themselves.

That last point deserves emphasis, because the choice of bracket engine
is not a detail.  We therefore also filled the same $d=1/3$ blocks with
the classical Talmi--Moshinsky formula~\cite{kamun-hob} as implemented
in the standard HOTB tool~\cite{hotb,hotb2}, which evaluates each matrix
element by an independent nested sum rather than a whole block at once,
in the two forms in which that code circulates: with a self-contained 6j
evaluated from raw binomials, and --- as the tool is distributed --- with
the 9j summed over the 6j of Tuzun, Burkhardt and Secrest
\cite{tuzun}, which looks its binomials up in a precomputed table and
exploits Regge symmetry.

The comparison separates two things easily confused.  Accuracy is
governed by the 6j, not by the formula: the two variants evaluate the
same expression, yet the table-based one agrees with the SU(3) engine to
$3\times10^{-14}$ at $E=12$ and still to $10^{-10}$ at $E=29$, where the
raw-binomial variant has lost every significant digit
($2\times10^{-1}$).  Agreement at the $10^{-10}$ level across that range,
between independent implementations of two different formulas, is a
strong mutual validation of both engines.  It is bounded not by
precision but by a fixed array: the distributed 6j dimensions its
binomial table at $n\le100$ and the range check guarding it is commented
out in the source, so from $E=30$ the lookup leaves the table and the
result is not a number at all.  Enlarging the table restores finite
values agreeing to $4\times10^{-11}$ at $E=30$, so this is a limit of
that implementation rather than of the formula --- but the tool as
distributed must not be used at $E\ge30$ without that change.

Speed, by contrast, is insensitive to the 6j: the table-based variant is
marginally slower, and both are two to three orders of magnitude slower
than the SU(3) engine ($41\times$ at $E=12$, $2077\times$ at $E=30$; a
factor $1321$ over all 481 blocks with $E\le30$), because each of the
$n^2$ elements is computed on its own.  A production code built on the
classical formula is therefore bracket-bound, and its end-to-end gain
from \eqref{eq:Ws} would be diluted well below the $1.8\times$ quoted
above --- while the $O(n^3)\to O(n^2)$ statement about the CFP-specific
step, which involves no bracket engine at all, is untouched.

Two further differences do not show up in the timings.  The eigenvalues
of the direct route are exact integers by construction, whereas the same
benchmark finds a deviation of $4.5\times10^{-6}$ of the computed
eigenvalues from the nearest integer over $E\le50$ (it is
$5.7\times10^{-8}$ if the sweep is stopped at $E=40$).  This figure is
identical for all three LAPACK drivers, which locates its origin: it is
the accuracy loss of Table~\ref{tab:fail40}, already present in the
bracket matrix handed to the eigensolver, and no choice of eigensolver
removes it.  And since
$\hat\Lambda$ has only four distinct eigenvalues, its eigenspaces are
massively degenerate, so a general eigensolver returns an arbitrary
orthonormal basis of each, which must then be re-analysed to recover the
symmetry class and the pair parity of every vector; the columns of
\eqref{eq:Ws} arrive already labelled, and each is supported on a single
$e_2$-parity block, so that a whole parity sector of every column of $K$
consists of exact structural zeros.

\section{Checking it against a code that knew nothing about it}
\label{sec:prod}

The strongest test of the construction is against data it had no hand in
producing.  We used the precomputed CFP matrices of the production
three-nucleon code \code{Trys\_kunai}, which computes $^3$H and $^3$He
bound states in a translationally invariant HO basis by the algebraic
approach of Ref.~\cite{fbs2024} (see also
Refs.~\cite{kamun-cfp,ncsm-ti}) and whose antisymmetrization input for
$J^\pi=\tfrac12^+$, $T=\tfrac12$ and $E=0,2,\dots,30$ was generated
years earlier by an independent numerical eigensolver of the
antisymmetrizer in the jj-coupled basis
\begin{equation}
  \bigl|\,[\,(s_1,\,(l_2\,(s_2s_3)S_2)\,J_2)\,J_{12},\;l_3\,]\,J;\;
  (t_1\,(t_2t_3)T_2)\,T\,\bigr\rangle ,
  \label{eq:prodbasis}
\end{equation}
with the pair $(e_2l_2)$, the relative coordinate $(e_3l_3)$, pair
antisymmetry $(-1)^{l_2+S_2+T_2}=-1$, and all shells $E'=e_2+e_3\le E$
stacked in one matrix; at $E=30$ this is $5712$ basis states and $1906$
physical columns.  We carried out three levels of comparison.

All 16 files parse, their stored columns are orthonormal to
$\sim5\times10^{-15}$, and an independent re-enumeration of the
production basis reproduces every row dimension $2,14,44,\dots,5712$.
These, like the physical dimensions below, are cumulative: the stored
matrix of a given $E$ stacks all shells $E'\le E$.

Next, the label counting of Sec.~\ref{sec:pseudo}, combined with the
spin--isospin content at $T=\tfrac12$
($[21]\otimes[21]=[3]\oplus[21]\oplus[111]$ for $S=\tfrac12$ and
$[3]\otimes[21]=[21]$ for $S=\tfrac32$) and the $L\otimes S\to
J=\tfrac12$ triangle rule, predicts the number of totally antisymmetric
states per shell,
\begin{equation}
  n_{\rm phys}(E')=\sum_{L}\Bigl[
  \bigl[\,n_{[3]}(E',L)+n_{[111]}(E',L)\,\bigr]\,\delta_{L\le1}
  +\tfrac12\,n_{[21]}(E',L)\,
   \bigl(\delta_{L\le1}+\delta_{1\le L\le2}\bigr)\Bigr] .
  \label{eq:nphys}
\end{equation}
This gives the per-shell values $1,4,10,19,30,\dots$, whose partial sums
$\sum_{E'\le E}n_{\rm phys}(E')$ reproduce the stored physical
dimensions $1,5,15,34,64,\dots,1906$ for every one of the 16 files,
with no diagonalization at all.

Finally we assembled the projected transposition $\hat P_{\rm eff}$, the
exchange of the spectator with a pair nucleon projected on the
pair-antisymmetric basis, in the production basis \eqref{eq:prodbasis}
from the present machinery: the orbital factor
$\hat\Pi_3\hat H_{1/3}\hat\Pi_3$ --- the analogue of \eqref{eq:ops} in the
basis \eqref{eq:prodbasis}, with $\hat\Pi_3\ket{i}=(-1)^{l_3}\ket{i}$ ---
computed by the same SU(3) engine, the spin and isospin exchange
operators, and the jj$\to$LS recoupling evaluated by explicit
magnetic-quantum-number sums.  For every shell $E'\le30$ its spectrum is
exactly the antisymmetrizer signature
$\{-1\ (\times\,n_{\rm phys}),\ +\tfrac12\ (\times\,n_{\rm spur})\}$,
equivalently $\hat A=\tfrac13(\Id-2\hat P_{\rm eff})\to\{1,0\}$, and
every stored production CFP $\ket{K_{\rm prod}}$ obeys
\begin{equation}
  \hat P_{\rm eff}\ket{K_{\rm prod}}=-\ket{K_{\rm prod}}
  \qquad\text{to }\lesssim10^{-10},
  \label{eq:opertest}
\end{equation}
after accounting for a single convention difference: the production code
couples the isospin in the order $(T_2t_1)T$, a diagonal phase
$(-1)^{T_2}$ per basis state relative to \eqref{eq:prodbasis}.  The
worst deviation over all 16 shells is $6.3\times10^{-10}$, at the
accuracy floor of the double-precision bracket engine.

As an end-to-end check we generated replacement CFP files from the
$(-1)$-eigenspace of the algebraically constructed $\hat P_{\rm eff}$, wrote
them in the production binary format, and substituted them into the full
production pipeline: the local position-space chiral interaction of
Saha \emph{et al.}~\cite{saha,chiralrev} at leading order with cutoffs
$(R_\pi,R_{ct})=(1.0,0.70)$~fm, together with the electromagnetic
terms, at $b=0.64$~fm.  The resulting
$^3$He ground-state energies are invariant
under the swap, as they must be for two orthonormal bases of the same
physical subspace: at $E=2$ they coincide with the original run to all
ten printed decimals, and at $E=10$ they agree to
$3\times10^{-10}$~MeV, exactly the level of the double-precision
subspace agreement (Table~\ref{tab:energy}).  The algebraic CFPs are
thus a drop-in replacement for the production antisymmetrization.

One limitation of this validation should be stated.  The production code
and the bracket engine \cite{su3hob} are both our own, so the tests
above establish consistency between two constructions that share an
institutional lineage, even though the stored CFP files were generated
years earlier by an independent numerical route.  The strongest
remaining check would be against the closed forms published for
$L=0$ and $L=1$ in the hyperspherical basis \cite{barnea}, or against
the projection construction of Ref.~\cite{amk1969}; both require a
change of basis that is a piece of work in its own right, and we have
not carried it out.

\begin{table}[htb]
\centering
\caption{Lowest $J^\pi=\tfrac12^+$, $T=\tfrac12$ eigenvalue returned by
the production driver when it is fed the original CFP files and when it
is fed the files generated by the present construction, in MeV, with the
full interaction --- nuclear plus electromagnetic --- in both cases.
The quantity of interest is the last column: the two runs differ from
each other by less than $10^{-9}$~MeV, which is all that is claimed
here.  The absolute values are reproduced from the driver as configured
for its own regression tests: the local position-space chiral
interaction of Saha \emph{et al.}~\cite{saha,chiralrev} at
\emph{leading order}, cutoffs $(R_\pi,R_{ct})=(1.0,0.70)$~fm,
$b=0.64$~fm, used bare --- the driver applies no model-space
renormalization, no similarity or Lee--Suzuki transformation.  They are
therefore not converged binding energies, both because $E=2$ and $E=10$
are far too small a model space and because a leading-order chiral
interaction is known to overbind; they are listed only so that the
comparison can be reproduced.}
\label{tab:energy}
\small
\begin{tabular}{ccc lll}
\toprule
$E$ & basis & physical & original CFPs & algebraic CFPs & difference\\
 & dim. & dim. & [MeV] & [MeV] & [MeV]\\
\midrule
 2 &  14 &   5 & $\phantom{-}15.9018188728$ & $\phantom{-}15.9018188728$
   & $0$\\
10 & 322 & 108 & $-7.5061794675$ & $-7.5061794678$
   & $3\times10^{-10}$\\
\bottomrule
\end{tabular}
\end{table}

\section{Conclusions}
\label{sec:concl}

We have shown that the coefficients of fractional parentage of three
identical particles in the oscillator Jacobi basis need not be computed
at all in the usual sense.  A single unitary, $\hat W=\hat H_1\hat\Phi$
--- the $d=1$ oscillator bracket together with the fixed
complexification built from $F=\dg(1,i)$ --- does two things at once.
It sends the class operator to the label swap times an integer,
$\hat W\hat\Lambda\hat W^{\dagger}=\sigma\hat J\hat Z$, and it carries
label parity into swap symmetry,
$\hat W\hat Q_s\hat W^{\dagger}=\hat R_s$.  Everything else follows
mechanically: the CFP is the single ket
$\ket{K_{i,s}}=\sqrt2\,\hat Q_s\hat W^{\dagger}\ket{i}$, its
orthonormality is the one line
$\braket{K_{i,s}}{K_{j,s'}}=\delta_{ss'}\me{i}{\Id+s\hat J}{j}$, and its
norm, $1+s\me{i}{\hat J}{i}$, absorbs into a single matrix element the
even/odd-$E$ case split that the original numerical experiments had to
treat separately.  Every CFP is therefore a single $\sqrt2$-scaled,
sign-decorated column of the $d=1$ HO bracket restricted to one
$e_2$-parity block, with an exact integer eigenvalue labelling its
S(3) class, and with the two members of a $[21]$ pair produced together
as a closed two-dimensional S(3) subspace.

The construction is uniform in $E$, $L$ and symmetry class and involves
no eigensolver of its own.  Our Fortran implementation passes the full
battery of checks --- including the class assignment, which is correct
in every block without exception --- for all 1301 blocks up to $E=50$,
far beyond the $E=23$ reach of the original numerical experiments.  The
residual errors are those of the bracket engine that supplies
$\hat H_1$, not of the recipe: in double precision they stay below
$10^{-8}$ throughout $E\le35$, and the quadruple-precision build
restores $10^{-22}$ or better in every block we recomputed beyond that.
Removing the eigensolver reduces the
CFP-specific work from $O(n^3)$ to $O(n^2)$, worth a factor $15$ at
$E=50$ against the fastest eigensolver we could configure, and
$1.74\times$ end to end once the shared cost of the brackets is
included.  Cross-validation against the independently
generated antisymmetrization data of a production three-nucleon code
confirms the construction state by state, and the $^3$He ground-state
energy of the full production pipeline is unchanged when the stored CFPs
are replaced by the algebraic ones.

Three directions follow naturally.  Since the cost and the accuracy are
now both dominated by the brackets themselves, further progress is a
question about the bracket engine rather than about antisymmetrization;
in particular, taking the $d=1$ brackets from a closed-form expression
\cite{trlifaj,kamun-hob} rather than from the SU(3) tower would remove
the last numerical eigensolver from the pipeline, and with it the
high-$E$ degradation that motivates the quadruple-precision build
altogether.  Second, the same $\hat W$ should be examined in the
hyperspherical basis, where the Raynal--Revai coefficients that
implement the corresponding rotation are themselves expressible through
Talmi--Moshinsky brackets \cite{novoselsky}: if the analogue of
\eqref{eq:intertwine} holds there, the recursive numerical construction
of hyperspherical CFPs would collapse in the same way.  Third, since the
complexification applies to a mirror in any single plane, the natural
next step is the four-particle case, where the corresponding operators
act on a Jacobi tree with more than one plane.

\section*{Acknowledgements}

The complexification idea at the heart of this construction --- to
surround the class operator with phases and let a single oscillator
bracket do the rest --- is due to Ramutis Kazys Kalinauskas, our friend
and colleague, who did not live to see it written down as a theorem.

\section*{Funding}

This research did not receive any specific grant from funding agencies
in the public, commercial, or not-for-profit sectors.

\section*{Declaration of competing interest}

The authors declare that they have no known competing financial
interests or personal relationships that could have appeared to
influence the work reported in this paper.

\end{document}